\documentclass[a4paper,11pt]{article}
\pdfoutput=1 % if your are submitting a pdflatex (i.e. if you have
\usepackage{jheppub} % for details on the use of the package, please
\usepackage[T1]{fontenc} % if needed
\usepackage{graphicx}  % needed for figures
\usepackage{dcolumn}   % needed for some tables
\usepackage{bm}        % for math
\usepackage{amssymb}   % for math
\usepackage{besphysics}
\usepackage{subfigure}
\usepackage{color}
\usepackage{xspace}
\usepackage{textpos}
\usepackage{epstopdf}
\usepackage{amsmath}
\usepackage{authblk}
\usepackage{epsfig}
\usepackage{multirow}
\usepackage{overpic}
\usepackage{colortbl}
\usepackage{booktabs}
\usepackage{algorithm, algorithmic}
\hypersetup{hidelinks}

\hypersetup{
colorlinks=true,
linkcolor=blue,
citecolor=blue,
urlcolor=blue
}

\leftline{Version 2 as of \today}

\title{Helicity amplitude analysis of $\chi_{cJ}\rightarrow\phi\phi$}

\collaborationImg{\includegraphics[width=0.15\textwidth, angle=90]{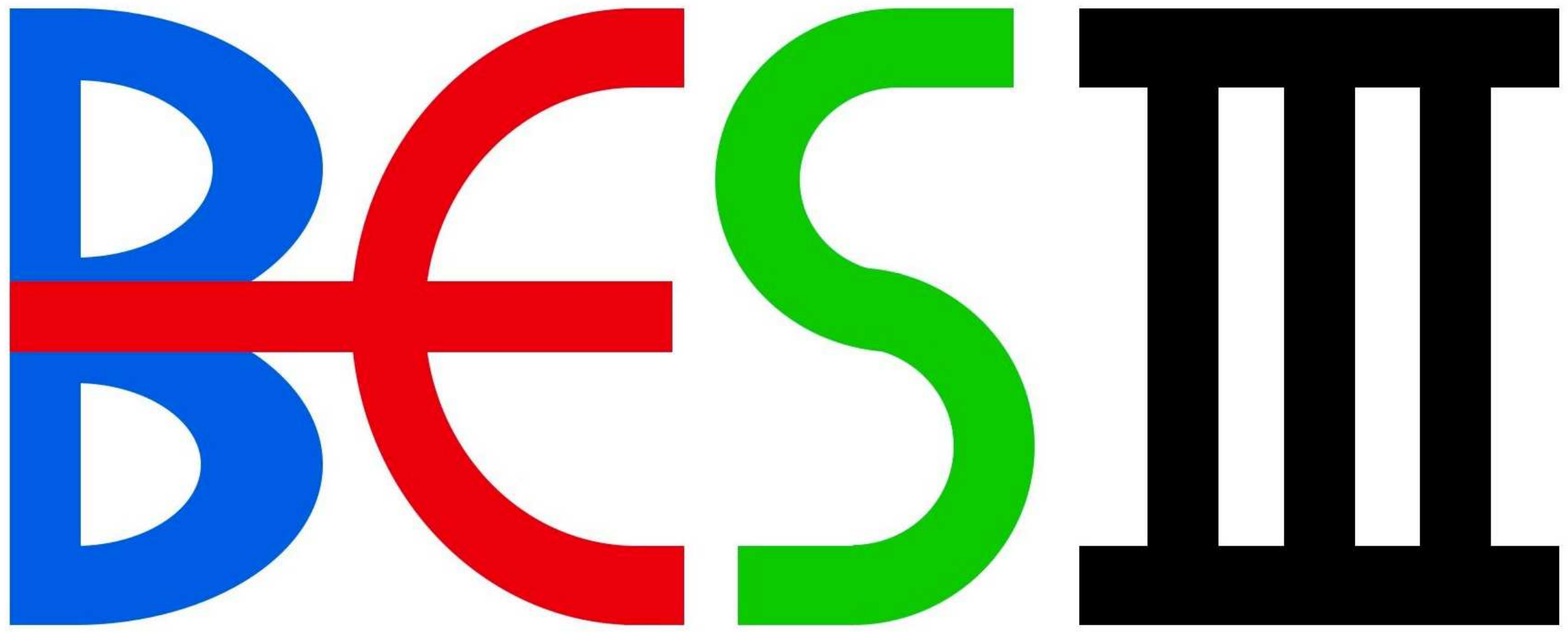}}
\collaboration{BESIII Collaboration}

\abstract{
Using (447.9 $\pm$ 2.3) million $\psip$ events collected with the BESIII detector, the decays of $\chi_{cJ} \rightarrow \phi\phi$ ($J$=0, 1, 2)  have been studied via the decay $\psi(3686)\rightarrow\gamma\chi_{cJ}$. The branching fractions of the decays $\chi_{cJ} \rightarrow \phi\phi$ ($J$=0, 1, 2) are determined to be $(8.48\pm0.26\pm0.27)\times10^{-4}$, $(4.36\pm0.13\pm0.18)\times10^{-4}$, and $(13.36\pm0.29\pm0.49)\times10^{-4}$, respectively, which are the most precise measurements to date. From a helicity amplitude analysis of the process $\psi(3686) \rightarrow \gamma \chi_{cJ}, \chi_{cJ}\rightarrow \phi\phi, \phi\rightarrow K^{+}K^{-}$, the polarization parameters of the $\chicJ\rightarrow\phiphi$ decays are determined for the first time. 
}

\begin{document}
%\linenumbers
\maketitle
\flushbottom

\section{\label{sec:level1}INTRODUCTION}
\vspace{-0.1cm}
Decays of the $\chi_{cJ} \  (J=0,1,2)$ states provide critical information to test Quantum Chromodynamics (QCD). In the quark model, the $\chi_{cJ}$ states are identified as $P$-wave triple charmonium states with spin, parity and charge conjugation $J^{++} \ (J=0,1,2)$. At leading order, the hadronic decays of $\chi_{cJ}$ are described by  annihilations of charm and anti-charm quarks into two gluons and subsequent production of light and/or strange quarks. Early theoretical calculations for exclusive decays of $\chi_{cJ}$ into light hadrons have yielded smaller branching fractions than experimental measurements~\cite{rf15.1,rf15.2,rf15.3}.

For a charmonium state $\psi(\lambda)$ decaying into light hadrons $h_1(\lambda_1)$ and $h_2(\lambda_2)$, the asymptotic behavior of the branching fraction is evaluated in perturbative QCD (pQCD) calculations~\cite{rfhel} as %\cite{rfhel,rfhel2},
\begin{equation}
\BR[\psi(\lambda) \rightarrow h_1(\lambda_1)h_2(\lambda_2)] \sim \left(\frac{\Lambda^2_{QCD}}{m^2_c}\right)^{| \lambda_1 + \lambda_2| + 2} ,
\end{equation}
where $\lambda, \lambda_1$, and $\lambda_2$ denote the helicities of the corresponding hadrons, $m_c\approx1.5$ GeV is the charm quark mass, and $\Lambda_{\rm{QCD}}$ denotes the $\mathrm{QCD}$ scale parameter. If the light-quark mass is neglected, the vector-gluon coupling conserves quark helicity leading to the helicity selection rule (HSR)~\cite{rf6}: $\lambda_1 + \lambda_2 = 0$. If the helicity configuration does not satisfy this relation, the branching fraction will be suppressed.

The $\chi_{cJ}\to\phi\phi$ decays clearly demonstrate that the decay mechanism of the $\chi_{cJ}$ particles is currently not well understood. Following pQCD calculations~\cite{rf3}, the $\chi_{c1}$ decay rate should be strongly suppressed compared to $\chi_{c0}$ and $\chi_{c2}$, due to HSR~\cite{rfhel} and the requirement of identical particle symmetry~\cite{rfyang}. However, BESIII reported similar branching fractions of $\chi_{cJ}\to\phi\phi$ decays for  $\chi_{c0}$, $\chi_{c1}$, and $\chi_{c2}$, namely $\BR(\chi_{c0}\to\phi\phi)=(7.8 \pm 0.4 \pm 0.8) \times 10^{-4}$, $\BR(\chi_{c1}\to\phi\phi)=(4.1\pm0.3\pm0.4)\times10^{-4}$, and $\BR(\chi_{c2}\to\phi\phi)=(10.7 \pm 0.4 \pm 1.1)\times10^{-4}$~\cite{rf14}.

The quark-pair creation model ($^3P_0$)~\cite{rf16} and  charm-loop ($D\bar D$ loop) contributions ~\cite{Huang:2021kfm,Liu:2009vv, Chen:2009ah} have been proposed to interpret the measured branching fractions with the model parameters obtained from data. In Ref.~\cite{Huang:2021kfm}, the analysis of the $\phi$ meson polarization is identified as a key measurement to probe hadronic-loop effects in the $\chi_{cJ} \to \phi \phi$ decays. Moreover, the ratios of the helicity amplitudes are found to be effective in the discrimination between the proposed models as these ratios are less sensitive to the parameters used in the evaluation of the model prediction. Table~\ref{pQCD_3P0} summarizes the helicity-amplitude ratios predicted by the considered theoretical models, where the uncertainties are due to the uncertainties of parameters involved in the calculation. The variable $x$ is defined as the ratio of transverse over the longitudinal polarized helicity amplitudes of the $\phi$ meson in $\chi_{c0}\to\phi\phi$: $x = \left|F^{0}_{1,1}/F^{0}_{0,0}\right|$ and the variables $\omega_{i}$ $(i=1,2,4)$ indicate the ratios of transverse over longitudinal polarized helicity amplitudes of the $\phi$ meson in $\chi_{c2}\to\phi\phi$: $\omega_1 = \left| F^{2}_{0,1}/F^{2}_{0,0}\right|$, $\omega_2 = \left| F^{2}_{1,-1}/F^{2}_{0,0}\right|$, $\omega_4 = \left| F^{2}_{1,1}/F^{2}_{0,0}\right|$, where $F^{J=0,2}_{\lambda_1,\lambda_2}$ are the helicity amplitudes. The $\chi_{c1}\to\phi\phi$ helicity amplitudes allow to test the validity of the identical particle symmetry: in this context the helicity-amplitude ratios  $u_1=|F^1_{1,0}/F^1_{0,1}|$ and $u_2=|F^1_{1,1}/F^1_{1,0}|$ are expected to be 1 and 0, respectively~\cite{rf16}.

\begin{table}[htbp]
    \caption{ \footnotesize  Numerical results of predictions from pQCD~\cite{rf3}, $^3P_0$~\cite{rf16} and $D\bar D$ loop models ~\cite{Huang:2021kfm}. \label{pQCD_3P0}}
    \centering
    %\vskip -0.2cm
    \begin{tabular}{lcccc}
    \toprule
    \hline
		Decay channel     &$\chi_{c0} \to \phi\phi$     &\multicolumn{3}{c}{$\chi_{c2} \to \phi\phi$}\\\hline
		Parameter         &$x$                        &$\omega_1$  &$\omega_2$  &$\omega_4$\\
		\midrule
		pQCD              &$0.293 \pm 0.030$            &$0.812 \pm 0.018$     &$1.647 \pm 0.067$      &$0.344 \pm 0.020$\\
		$^3P_0$           &$0.515 \pm 0.029$            &$1.399 \pm 0.580$     &$0.971 \pm 0.275$      &$0.406 \pm 0.017$\\
        $D\bar{D}$ loop   &$0.359\pm0.019$  &$1.285\pm0.017$ &$5.110\pm0.057$ &$0.465\pm0.002$\\
    \hline
    \bottomrule
    \end{tabular}
\end{table}

In this analysis, the $\chi_{cJ}\to\phi\phi$ decays are studied to extract the polarization parameters from a data sample corresponding to 448.1 million $\psi(3686)$ events~\cite{numofpsi} collected in $e^{+}e^{-}$ annihilation with the BESIII detector. The measurements of polarization parameters provide further information to understand $\chi_{cJ}$ decay mechanisms, and to test the quark-pair-creation model and charm-quark loop contributions to reveal the evasion of HSR~\cite{Liu:2009vv} in the $\chicJ$ decays. Moreover, improved measurements of the $\chi_{cJ}\to\phi\phi$ branching fractions are reported.

%=================
\section{\label{sec:level2}DETECTOR AND MONTE CARLO SIMULATION}
The BESIII detector~\cite{bes3} records symmetric $e^+e^-$ collisions
provided by the BEPCII storage ring~\cite{bepc2}, which operates
in the center-of-mass (CM) energy range from 2.00 to 4.95~GeV, with a peak luminosity of $1\times 10^{33}~\mathrm{cm}^{-2}\mathrm{s}^{-1}$ achieved at $\sqrt{s} = 3.77$ GeV.
The cylindrical core of the BESIII detector covers 93\% of the full solid angle and consists of a helium-based multilayer drift chamber~(MDC), a plastic scintillator time-of-flight
system~(TOF), and a CsI(Tl) electromagnetic calorimeter~(EMC), which are all enclosed in a superconducting solenoidal magnet providing a 1.0~T~\cite{visualization} (0.9~T in 2012) magnetic field. The solenoid is supported by an octagonal flux-return yoke with resistive plate counter muon identification modules interleaved with steel.
The charged-particle momentum resolution at $1~{\rm GeV}/c$ is $0.5\%$, and the ${\rm d}E/{\rm d}x$ resolution is $6\%$ for electrons from Bhabha scattering. The EMC measures photon energies with a resolution of $2.5\%$ ($5\%$) at $1$~GeV in the barrel (end cap) region. The time resolution in the TOF barrel region is 68~ps, while that in the end cap region is 110~ps.

Simulated data samples produced with a {\sc geant4}-based~\cite{geant4} Monte Carlo (MC) package, which includes the geometric description of the BESIII detector and the detector response, are used to determine detection efficiencies and to estimate backgrounds. The simulation models the beam energy spread and initial state radiation (ISR) in the $e^+e^-$ annihilations with the generator {\sc kkmc}~\cite{Jadach:2000eu,Jadach:1999vf}. Signal MC events for $\chi_{cJ}(J=0,1,2)\to\phi\phi$ are generated by using the amplitude model with helicity amplitude ratios fixed to the results of this amplitude analysis. An inclusive MC sample, which includes the production of the 506 millon $\psi(3686)$ resonance, the ISR production of the $J/\psi$, and the continuum processes incorporated in {\sc kkmc}, is used for studying background contributions. All particle decays are modelled with {\sc
evtgen}~\cite{Lange:2001uf,Ping:2008zz} using branching fractions either taken from the Particle Data Group~\cite{rf12}, when available, or otherwise estimated with {\sc lundcharm}~\cite{Chen:2000tv,Yang:2014vra}. Final state radiation~(FSR)
from charged final state particles is incorporated using the {\sc photos} package~\cite{Richter-Was:1992hxq}.

%===========================================
\section{\label{sec:level3}EVENT SELECTION AND BACKGROUND ANALYSIS}

The $\chicJ\to\phi\phi$ candidates are selected from the process $e^{+}e^{-}\to\psi(3686)\to\gamma\chi_{cJ}$. The $\gamma K^{+}K^{-}K^{+}K^{-}$ final state is selected by requiring four charged tracks with zero net charge and at least one photon shower. Charged tracks detected in the MDC are required to be within a polar angle ($\theta$) range of $|\rm{cos\theta}|<0.93$, where $\theta$ is defined with respect to the $z$-axis, which is the symmetry axis of the MDC. For charged tracks not originating from $K_S^0$ or $\Lambda$ decays, the distance of closest approach to the interaction point (IP) must be less than 10\,cm along the $z$-axis, $|V_{z}|$,  
and less than 1\,cm in the transverse plane, $|V_{xy}|$. Photon candidates are identified using showers in the EMC. The deposited energy of each shower must be more than 25~MeV in the barrel region ($|\cos \theta|< 0.80$) and more than 50~MeV in the end cap region ($0.86 <|\cos \theta|< 0.92$). To exclude showers that originate from charged tracks, the angle subtended by the EMC shower and the position of the closest charged track at the EMC must be greater than 10 degrees as measured from the IP. To suppress electronic noise and showers unrelated to the event, the difference between the EMC time and the event start time is required to be within [0,700]\,ns.

\begin{figure}[htbp]
\centering
\mbox{
    \begin{overpic}[scale=0.36]{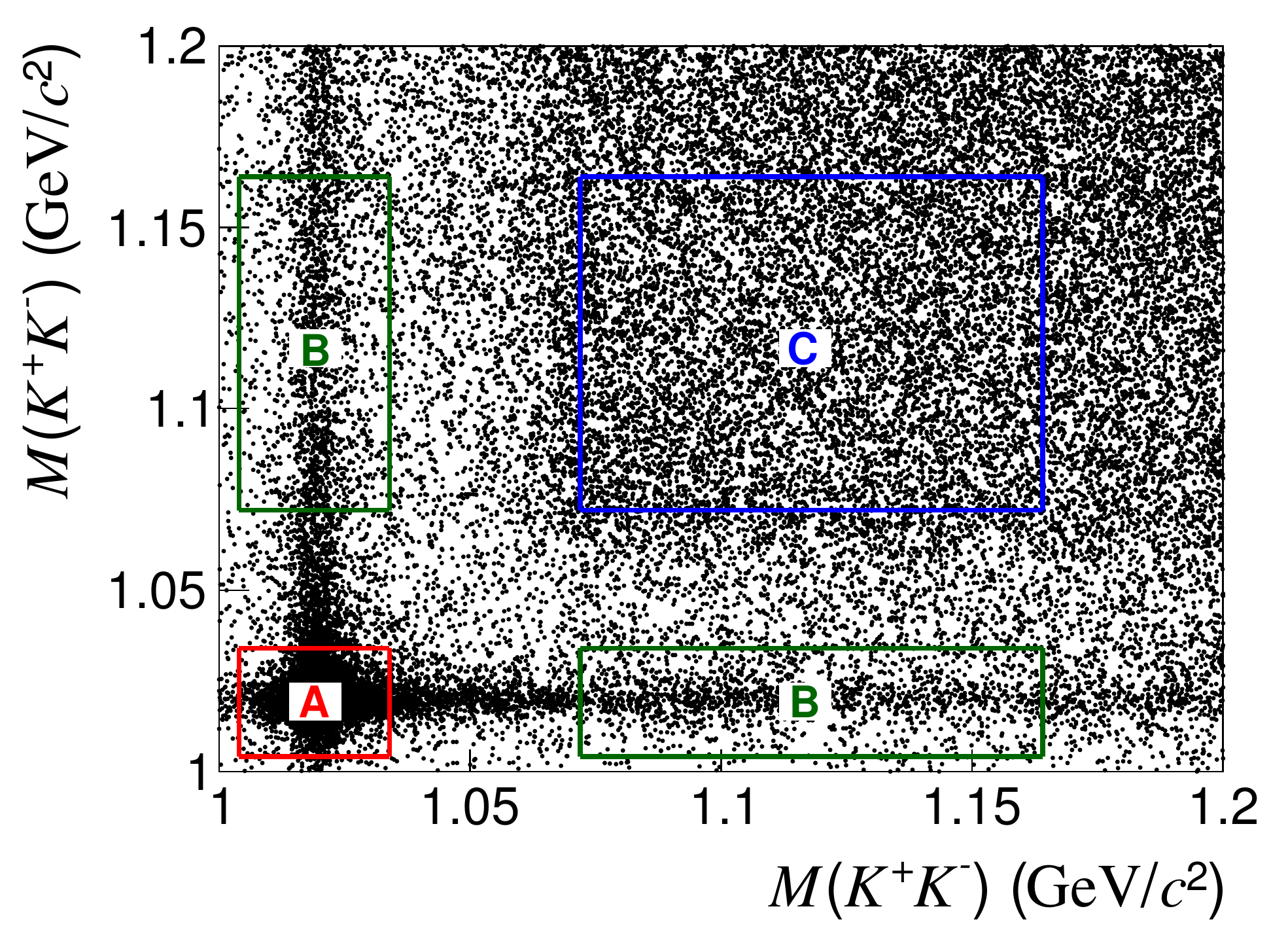} \end{overpic}
    \begin{overpic}[scale=0.36]{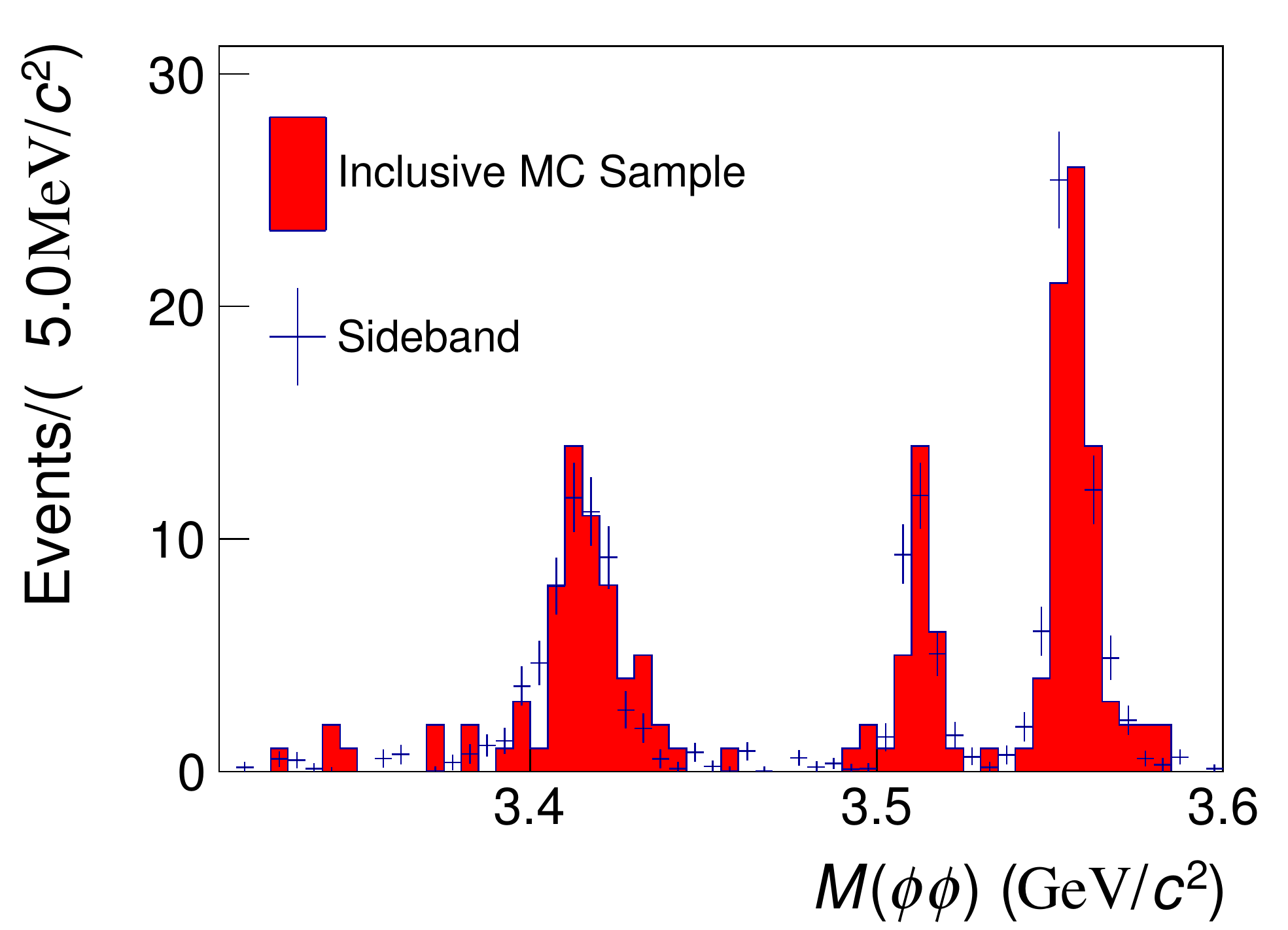} \end{overpic}
 }
 {\caption{left: Distribution of $M(K^+K^-)$ versus $M(K^+K^-)$, two pairs of $K^{+}K^{-}$ are plotted symmetrically; right: $\chicj$ peaking backgrounds ($\chi_{cJ}\to\phi K^{+}K^{-}$ and  $K^{+}K^{-}K^{+}K^{-}$) estimated from inclusive MC events and the sideband events in data.  \label{kksact}}}
\end{figure}

To improve the momentum resolution and to reduce the background contributions, the four charged tracks that are assumed to be kaons and one photon candidate are subjected to a kinematic fit constrained by four-momentum conservation (4C), with the hypothesis of the candidate events coming from $\psip\to\gamma2( K^{+}K^{-})$. Among all the photon candidates reconstructed in the event, only that providing the least $\chi^2_{\rm{4C}}$ value in the kinematic fit ($\chi^2_{\rm{4C}}$) is retained. Signal candidates are selected by requiring $\chi^{2}_{\rm{4C}}<60$, which is optimized by the figure of merit $\frac{S}{\sqrt{S+B}}$, where $S$ and $B$ are the normalized numbers of signal and background events obtained from signal and inclusive MC sample, respectively. The intermediate $\phi$ mesons are reconstructed as the $K^{+}K^{-}$  combinations that minimize the discriminator $\delta = \sqrt{(M_{K^{+}K^{-}}^{(1)} - m_{\phi})^2 +  (M_{K^{+}K^{-}}^{(2)} - m_{\phi})^2 }$, where $M_{K^{+}K^{-}}^{(i)}$ is the invariant mass of the $K^{+}K^{-}$ combination $i$ and $m_{\phi}$ is the known $\phi$ mass~\cite{rf12}. The $\phi$ signal region is defined as $|M_{K^{+}K^{-}} - m_{\phi}|< 0.015$ GeV/$c^{2}$ from the study of simulated events, as shown in the area $A$ in Fig.~\ref{kksact}. This mass window is also applied for signal MC samples to obtain efficiency. Background events are subtracted using two sideband regions of data events corresponding to $\chi_{cJ}\to\phi K^{+}K^{-}$ and  $K^{+}K^{-}K^{+}K^{-}$ in the following amplitude analysis. The $\phi K^{+}K^{-}$ sideband region requires one pair of kaons to satisfy $|M_{K^{+}K^{-}}^{(1)} - m_{\phi}-0.1~{\rm GeV}/c^{2} |< 3\times 0.015$ GeV/$c^{2}$ and the other pair to satisfy $|M_{K^{+}K^{-}}^{(2)} - m_{\phi}|< 0.015$ GeV/$c^{2}$, illustrated as the area $B$ in Fig. \ref{kksact}. The $K^{+}K^{-}K^{+}K^{-}$ sideband region requires both pairs of kaons to satisfy  $|M_{K^{+}K^{-}}^{(1,2)} - m_{\phi} - 0.1~{\rm GeV/}c^{2}|< 3\times 0.015$ GeV/$c^{2}$, illustrated as the area $C$ in Fig.~\ref{kksact}.

%======================================
%======================================

By analyzing an inclusive MC sample corresponding to 506 million $\psi(3686)$ events, the number of background events is estimated to be 153 after applying the same reconstruction and selection procedure as for real data, which is about 1.7$\%$ in total MC events. The decay with the same final state, $\chicJ\to K^{+}K^{-}K^{+}K^{-}$ without intermediate states, is found to be the dominant background according the analysis of the event type of inclusive MC~\cite{topology}.
The shape and amount of the background events in the signal region are estimated using the events in the sideband regions $B$ and $C$. The scale factor for the events in each sideband region is determined through exclusive simulations of $\chi_{cJ}$ decays into $\phi K^{+}K^{-}$ and $K^{+}K^{-}K^{+}K^{-}$. To be more specific, the number of selected sideband events is determined by the ratio of the number of events in the sideband region to the number of events in the signal region with an MC simulation for the $\phi K^{+}K^{-}$ and $K^{+}K^{-}K^{+}K^{-}$ decays. The background contributions evaluated with the inclusive MC sample and with the sideband events in data are in good agreement as shown in the right of Fig.~\ref{kksact}.

The background contribution from the continuum process $e^{+}e^{-} \to 2(K^{+}K^{-})\gamma$ is studied with a data sample of $e^{+}e^{-}$ annihilations collected at a center-of-mass (CM) system energy of 3.65 $\textrm{GeV}$, just below the $\psi(3686)$ resonance, and with an integrated luminosity of 42.6 pb$^{-1}$. The signal selection previously described, with the kinematic fit constraint adapted to the different CM system, does not show any event in the $\chi_{cJ}$ signal region. The contribution of the continuum process is therefore negligible.

%==============================================================
\section{AMPLITUDE ANALYSIS}
\subsection{Helicity System}
\begin{figure}[htbp]
\setlength{\abovecaptionskip}{-0.6 cm}
\setlength{\belowcaptionskip}{0.cm}
\centering
\mbox{
 \begin{overpic}[scale=0.55]{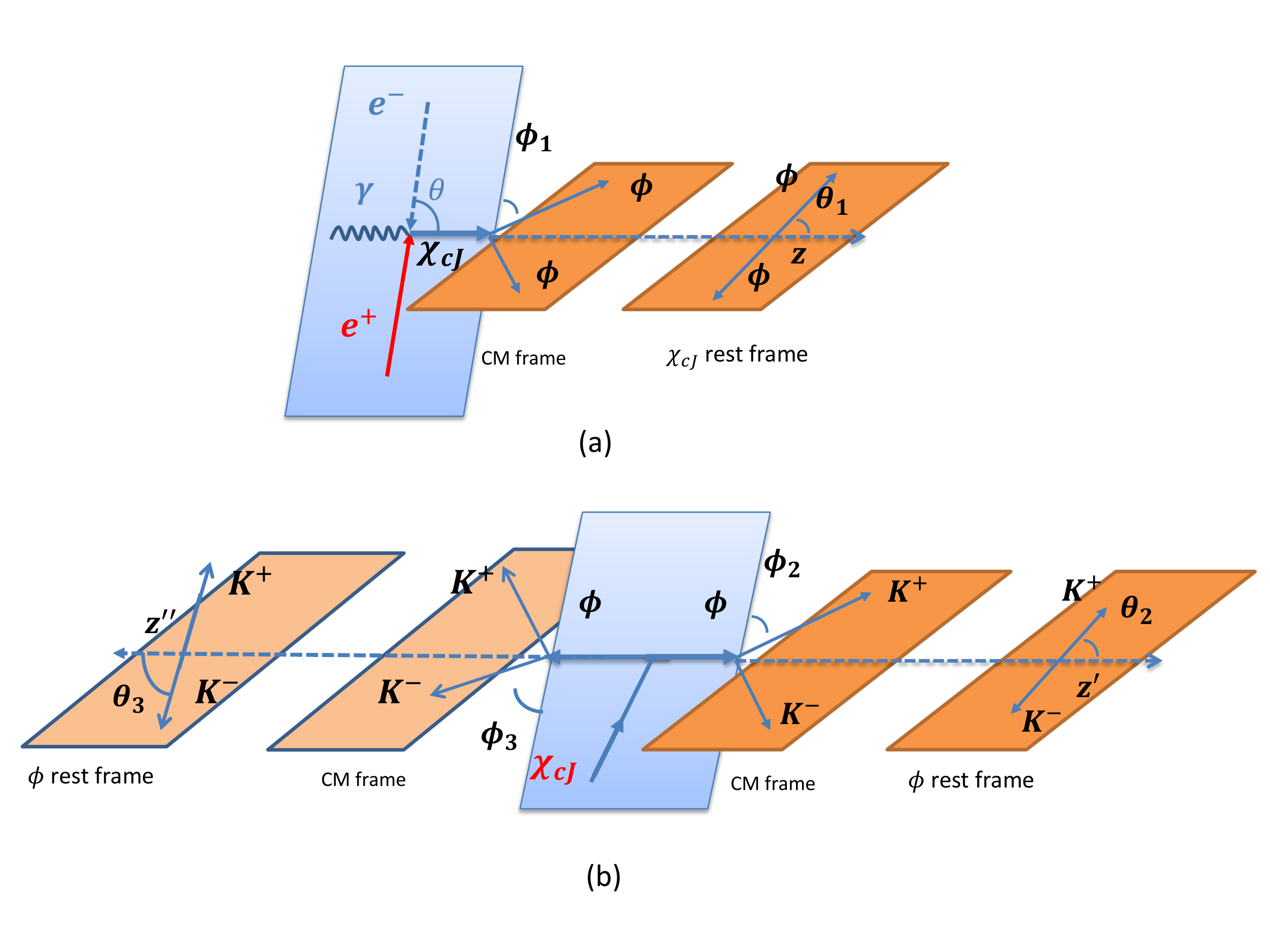} \end{overpic}
 }
 {\caption{Definitions of helicity angles.}\label{helsys}}
\end{figure}

The selected $\chi_{cJ}\to\phi\phi$ events are subject of a helicity amplitude analysis to determine the polarization parameters. In the amplitude analysis, the joint angular distribution for the sequential decays $\ee\to\psi(3686)\to\gamma\chicJ,~\chicJ\to\phiphi$ and $\phi\to\kk$ is constructed in the helicity system of various intermediate resonances in the process, as shown in Fig.~\ref{helsys}. The helicity axis for a particular decay is characterized by the momentum direction of the decaying particle in its mother particle rest frame. Starting from the unpolarized $e^{+}e^{-}$ initial system, the helicity axes are defined as described below.
\begin{itemize}
\item For $\ee\to\psip\to\gamma\chicJ$, the polar angle $\theta_0$ is defined as the angle spanned between the $\chicJ$ moving direction and the $e^+$ beam direction, which corresponds to the polar angle of the $\chicJ$ momentum in the $\ee$ CM system. The corresponding azimuthal angle follows Fig.~\ref{helsys} and cancels out in the amplitude of this decay.
\item For $\chicJ\to\phiphi$, the momenta of the two $\phi$ mesons define the $\chicJ$ decay plane. Due to momentum conservation, the momenta of the $\chicJ$ meson and the two $\phi$ mesons have to be located in the same decay plane. After boosting the two $\phi$ momenta to the $\chicJ$ rest frame, they are still in the same decay plane. Then the polar angle $\theta_1$ is defined as the angle between the $\chicJ$ momentum and the $\phi$ momentum in the $\chicJ$ rest frame. The azimuthal angle $\phi_1$ is defined as the angle between the $\chicJ$ production and decay planes.
\item For each of the two $\phi\to \kk$ decays, the polar angle $\theta_2$($\theta_3$) is defined as the angle between the $K^+$($K^-$) momentum in the $\phi$ rest frame and the $\phi$ momentum in the $\chicJ$ rest frame. The corresponding angles $\phi_2$ and $\phi_3$ are defined as the angle spanned between the $\phi$ production and decay planes.
\end{itemize}
%======================================
%======================================

\subsection{Helicity Amplitude}

Let $X$ be a resonance of spin-parity $J^{\eta}$ (with $z$-component $M$) and mass $\mu$ which decays into two particles 1 and 2,  $X \rightarrow 1 + 2$, where the particle $i$ has spin $S_{i}$, intrinsic parity $\eta_{i}$, and helicity $\lambda_{i}$ (with $i=1,2$). In the rest frame of the resonance $X$, $\vec{p}$ represents the momentum of the particle 1 with the spherical angles given by $\Omega$ = ($\theta$, $\phi$). $\mathcal{M}$ is the decay operator. Then the helicity amplitude $A$ for the two-body decay $X \rightarrow 1 + 2$ can be written as~\cite{chung.1,chung.2,chung.3}

\begin{equation*}
\begin{aligned}
A &= \left \langle \vec{p}\lambda_{1}; - \vec{p}\lambda_{2} \left | \mathcal{M}\right | JM \right \rangle \\
&=4\pi\left(\frac{\mu}{p}\right)^{\frac{1}{2}} \left \langle \phi\theta\lambda_{1}\lambda_{2}|JM\lambda_{1}\lambda_{2} \right \rangle\left \langle JM\lambda_{1}\lambda_{2} \left | \mathcal{M}\right |JM\right \rangle\\
&=N_{J} F_{\lambda_{1},\lambda_{2}}^{J}D_{M,\lambda}^{J*}(\phi,\theta,0), \lambda = \lambda_{1}-\lambda_{2},
\end{aligned}
\end{equation*}
where $N_J=4\pi$ $(\frac{\mu}{p})^{\frac{1}{2}}$ is a normalization factor and $D_{M,\lambda}^{J*}(\phi,\theta,0)$ is the $D$-function~\cite{Chung:1971ri}. Generally, the helicity amplitudes $ F_{\lambda_{1},\lambda_{2}}^{J}$ depend on the momenta of the final state particles. It can be expanded in terms of the $L$-$S$ coupling scheme~\cite{chung.1,vf}:

\begin{equation}
\label{helamp}
 F_{\lambda_{1}, \lambda_{2}}=\sum\limits_{LS} g_{LS}\sqrt{\frac{2L+1}{2J+1}}
 \langle L0S\lambda|J\lambda\rangle
 \langle \eta_{1}\lambda_{1}\eta_{2}-\lambda_{2}|S\lambda\rangle r^{L}B_{L}(r)/B_{L}(r_0),
\end{equation}
where $g_{LS}$ is the coupling constant in the $L$-$S$ coupling
scheme; $\langle L0S\lambda|J\lambda\rangle$ and $ \langle \eta_{1}\lambda_{1}\eta_{2}-\lambda_{2}|S\lambda\rangle$  denote the Clebsch-Gordan coefficients; $r$ is the momentum of the two final state particles, measured in the resonance rest frame, and $r_0$ 
is the corresponding quantity evaluated at the nominal mass of the resonance; $B_L$ is a barrier factor, which depends on the angular momenta $L$ reaching from 0 up to 4 and can be written as~\cite{barrierform}
\begin{eqnarray}
B_0(r)/B_0(r_0)&=&1,\nonumber\\
B_1(r)/B_1(r_0)&=&\frac{\sqrt{1+(dr_0)^{2}}}{\sqrt{1+(dr)^{2}}},\nonumber\\
B_2(r)/B_2(r_0)&=&\frac{\sqrt{9+3(dr_0)^{2}+(dr_0)^{4}}}{\sqrt{9+3(dr)^{2}+(dr)^{4}}},\\
B_3(r)/B_3(r_0)&=&\frac{\sqrt{225+45(dr_0)^{2}+6(dr_0)^{4}+(dr_0)^{6}}}{\sqrt{225+45(dr)^{2}+6(dr)^{4}+(dr)^{6}}},\nonumber\\
B_4(r)/B_4(r_0)&=&\frac{\sqrt{11025+1575(dr_0)^{2}+135(dr_0)^{4}+10(dr_0)^{6}+(dr_0)^{8}}}{\sqrt{11025+1575(dr)^{2}+135(dr)^{4}+10(dr)^{6}+(dr)^{8}}},\nonumber
\end{eqnarray}
where $d$= 3 GeV$^{-1}$ is a constant~\cite{lhcb}.

Table~\ref{tab helangle} shows the definitions of helicity angles and amplitudes for the sequential process $\psip$ $\rightarrow$ $\gamma$ $R_{i}$, $R_{i}$ $\rightarrow$ $\phi\phi$, and $\phi$ $\rightarrow$ $K^{+}K^{-}$.

\begin{table}[htbp]
%    \vskip -0.2cm
%    \hskip -0.5 cm
\begin{center}
\caption{ \label{tab helangle} \footnotesize Definitions of helicity angles and amplitudes of sequential decays.}
    \vskip 0.2cm
    \hskip 0.3 cm
\begin{normalsize}
\begin{tabular}{ m{5cm}<{\centering} | m{2.5cm}<{\centering} | m{4.5cm}<{\centering}}
\hline
Decay Mode & Helicity Angle & Amplitude \\
\hline
$\psip$($M$)  $\rightarrow$ $R_{i}(\lambda_{R})$ $\gamma$($\lambda_{\gamma}$) & $\theta_{0}$ & $A_{\lambda_{\gamma},\lambda_{R}}^{1}$ $D_{M,\lambda_{R}-\lambda_{\gamma}}^{1*}$(0,$\theta_{0}$,0)\\
%\hline
$R_{i}(\lambda_{R})$ $\rightarrow$ $\phi(\lambda_{1}) \phi(\lambda_{2})$ & $\theta_{1}$,$\phi_{1}$ & $F_{\lambda_{1},\lambda_{2}}^{J}$ $D_{\lambda_{R},\lambda_{1}-\lambda_{2}}^{J*}$($\phi_{1}$,$\theta_{0}$,0)\\
%\hline
$\phi(\lambda_{1})$ $\rightarrow$ $K^{+}(0^{-})$ $K^{-}(0^{-})$ & $\theta_{2}$,$\phi_{2}$ & $B_{0,0}^{1}$ $D_{\lambda_{1},0}^{1*}$($\phi_{2}$,$\theta_{2}$,0)  \\
%\hline
$\phi(\lambda_{2})$ $\rightarrow$ $K^{+}(0^{-})$ $K^{-}(0^{-})$ & $\theta_{3}$,$\phi_{3}$ & $B_{0,0}^{1}$ $D_{\lambda_{2},0}^{1*}$($\phi_{3}$,$\theta_{3}$,0)  \\
\hline
\end{tabular}
\end{normalsize}
\end{center}
\end{table}

Then the joint amplitude for the sequential process is obtained by
\begin{eqnarray}\label{ampform}
\mathcal{M}(R_i)  &=& {1\over 2}\sum_{M,\lambda_{R},\lambda_{1},\lambda_{2}}
 A_{\lambda_{R},\lambda_{\gamma}}^{1}D_{M,\lambda_{R}-\lambda_{\gamma}}^{1*}(0,\theta_{0},0) F_{\lambda_{1},\lambda_{2}}^{J}D_{\lambda_{R},\lambda_{1}-\lambda_{2}}^{J*}(\phi_{1},\theta_{0},0)\nonumber\\
 &\times&B_{0,0}^{1}D_{\lambda_{1},0}^{1*}(\phi_{2},\theta_{2},0)B_{0,0}^{1}D_{\lambda_{2},0}^{1*}(\phi_{3},\theta_{3},0)BW(m_{\phi\phi},m_{i},\Gamma_{i}),
\end{eqnarray}
with
\begin{eqnarray}\label{BWform}
BW(m_{\phi\phi},m_i, \Gamma_i) = {1 \over m_{\phi\phi}^2-m_i^2 + i m_i \Gamma_i },
\end{eqnarray}
where $R_{i}$ can assume a resonant $\chi_{cJ}$ or a non-resonant (NR) contribution; $M$ is the $z$-projection of the $\psip$ spin with $M=\pm1$ since it is produced from unpolarized $\ee$ beams; $A_{\lambda_{R},\lambda_{\gamma}},~F_{\lambda_{1},\lambda_{2}}^J$, and $B_{0,0}^{1}$ are helicity amplitudes which are expanded in terms of the $L$-$S$ coupling constant $g_{LS}$ according to Eq.~(\ref{helamp}), $m_{\phi\phi}$ is the invariant mass of the $\phi$ meson pair, $m_{i}$ and $\Gamma_{i}$ are mass and width of the corresponding $\chi_{cJ}$, and $BW(m_{\phi\phi},m_{i},\Gamma_{i})$ is a Breit-Wigner (BW) function for $\chicJ$. In a consistent description of the non-resonant contribution, NR $\rightarrow$ $\phi$ $\phi$, in which the $\phi\phi$ system has quantum numbers $J^{P}=0^\pm,1^+$ and $2^+$, the BW function is set to 1. The BW functions for the two $\phi$ mesons are isolated from the above equation. They are taken into account in the MC event generation and used to estimate the normalization factor for the likelihood function. The $\phi$ BW function and the mass resolution are well simulated in the phase-space MC events of the decay $\psip\to \gamma\phi\phi\to\gamma2\kk$. It is noteworthy that this partial wave expansion ensures parity conservation of the helicity amplitude in the sequential decays. The allowed values for $L$ and $S$ of the corresponding subprocesses are given in Table~\ref{tab couple}. The $g_{LS}$ parameters involved in the fit are taken as complex numbers, and their values are determined from the fit to data.
\begin{table}[htbp]
\begin{center}
\caption{ \label{tab couple} \footnotesize Involved partial waves $(LS)$ in the joint angular distribution.}
    \vskip 0.6cm
    \hskip 0.5 cm
\begin{normalsize}
\begin{tabular}{ m{6cm}<{\centering} | m{6cm}<{\centering} }
\hline
Decay & Partial waves $(LS)$ \\
\hline
$\psip \rightarrow \gamma \chi_{c0}$ & (01) , (21)\\
$\psip  \rightarrow \gamma \chi_{c1}$ &  (01) , (21) , (22)\\
$\psip  \rightarrow \gamma \chi_{c2}$ &  (01) , (21) , (22) , (23) , (43)\\
$\chi_{c0}$ or NR$(0^{+}) \rightarrow\phi \phi$ &  (00) , (22)\\
$\chi_{c1}$ or NR$(1^{+}) \rightarrow\phi \phi$ &  (01) , (21) , (22)\\
$\chi_{c2}$ or NR$(2^{+}) \rightarrow\phi \phi$ &  (02) , (20) , (21) , (22) , (42)\\
NR($0^{-}) \rightarrow\phi \phi$ &  (11) \\
$\phi  \rightarrow K^{+}K^{-}$ & (10) \\
\hline
\end{tabular}
\end{normalsize}
\end{center}
\end{table}

The partial decay rate of $\psi(3686)$ is given by
\begin{equation}
\mathrm{d}\sigma \propto \frac{1}{2}{\sum}_{M,\lambda_{\gamma}}\left|\sum_{R_i}\mathcal{M}(R_i)\right|^2\mathrm{d}\Phi,
\end{equation}
where $\mathrm{d}\Phi$ is the standard phase space for the decay $\psip\to\gamma\phi\phi$ with $\phi\to\kk$. The summation is taken over $M=\pm1$ due to the fact that the $\psip$ is produced from unpolarized $\ee$ beams.

%==============================================
\subsection{$\chi_{cJ}$ Mass Resolution}
The widths of the $\chi_{c0}$, $\chi_{c1}$, and $\chi_{c2}$ states are $\Gamma= 10.8\pm0.6, 0.84\pm0.04, \rm{and}~1.97\pm0.09$ MeV~\cite{rf12}, respectively. Meanwhile, the mass resolution, whose magnitude is 4.6 MeV, is greater than $\chi_{c1}$ and $\chi_{c2}$ widths and therefore crucial in this analysis. The observed resonances can be well approximated using $\chicJ$ lineshapes convolved with the mass resolution function. Therefore, in the amplitude analysis, the BW function for the $\chi_{cJ}$ in Eq.~(\ref{ampform}) is replaced with
\begin{equation}
|BW(m_{\phi\phi},m_{0},\Gamma)|^2 \to\int_{-\infty}^{+\infty} |BW(m'_{\phi\phi},m_{0},\Gamma)|^2 R(m'_{\phi\phi}-m_{\phi\phi})\mathrm{d}m'_{\phi\phi},
\label{res}
\end{equation}
where $R(m)$ is the mass resolution function, which is determined from MC simulations. The detector resolution function is parametrized with a three-Gaussian function, i.e.
\begin{equation}
\begin{aligned}
		R(m_{\phiphi}) = f_{1}|G(m_{\phi\phi},m_{1},\sigma_{1})|^2 + f_{2}|G(m_{\phi\phi},m_{2},\sigma_{2})|^2 + (1-f_{1}-f_{2})|G(m_{\phi\phi},m_{3},\sigma_{3})|^2,
\end{aligned}
\end{equation}
where $m_j$ and $\sigma_j$ are the mean and width of the $j$-th Gaussian function ($G$), $j=1,2,3$. For each $\chi_{cJ}$ state, the fraction parameters $f_1$ and $f_2$ are determined by fitting the $\chi_{cJ}$ lineshape.

%==========================================================
\subsection{Fit Method}
The relative magnitudes and phases of the coupling constants, $g_{LS}$, are determined
by an unbinned maximum likelihood fit. The joint likelihood
for observing $N$ events in the data set is
\begin{equation}
\mathcal{L}=\prod_{i=1}^N P(x_i),
\end{equation}
where $P(x_i)$ is the probability to produce event $i$ with a set of four-momenta $x_i$. The normalized $P(x_i)$ is calculated from the differential cross section
\begin{equation}\label{CrossSectionFormula}
P(x_i)={(\mathrm{d}\sigma /\mathrm{d}\Phi)_i \over \sigma_{\rm MC}},
\end{equation}
where $\sigma_{\mathrm MC}$ is calculated using a set of $\mathrm{MC}$ events. $\mathrm{MC}$ events are generated according to a phase space distribution and are subject to the detector simulation. Simulated events pass the same reconstruction and selection criteria as events recorded by the detector. For an MC sample with sufficient statistics,
$\sigma_{\rm{MC}}$ is evaluated with
\begin{equation}\label{normfct}
\sigma_{\rm MC}={1\over N_{\rm MC}}\sum_{i=1}^{N_{\rm MC}}\left({\mathrm{d}\sigma\over \mathrm{d}\Phi}\right)_i,
\end{equation}
where $N_{\rm MC}$ is the number of events passing all selection criteria. For technical reasons, rather than maximizing $\mathcal{L}$, the object function, $S=-\ln\mathcal{L}$, is minimized using the package MINUIT \cite {minuitRef}. To subtract background events, the likelihood function is calculated for both data ($\ln\mathcal{L}_\textrm{data}$) and background selected from the sideband regions ($\ln\mathcal{L}_\textrm{bkg}$), i.e.

\begin{equation}\label{objectfun}
S=-\ln\mathcal{L} = -\ln\mathcal{L}_\textrm{data}+\ln\mathcal{L}_\textrm{bkg}.
\end{equation}

With the parameters obtained from the fit, the signal yield of a given resonance can be estimated as
\begin{equation}\label{yieldsFormula}
N_i=\frac{\sigma_i(N_\textrm{obs}-N_\textrm{bkg})}{\sigma_{tot}},
\end{equation}
where $\sigma_i$ is evaluated with the differential cross section for the $i$-th resonance, $\sigma_{tot}$ is the total cross section including interference effects, $N_\textrm{obs}=8664$ is the number of observed events, and $N_\textrm{bkg}=166$ is the number of background events.

The statistical uncertainty $\delta N_i$ associated with the signal yield $N_i$ is estimated according to the error propagation formula using the covariance matrix $V$ which is obtained in the MIGRAD fit,
\begin{equation}\label{staterr}
\delta N_{i}^{2} = \sum_{m=1}^{N_\textrm{pars}}\sum_{n=1}^{N_\textrm{pars}}\left({\partial N_i\over \partial X_m}{\partial N_i\over \partial X_n}\right)_{X={\bf \mu}}V_{mn}(X),
\end{equation}
where $X$ is the vector parameters, and ${\bf \mu}$ contains the fitted values for all parameters. The sum runs over all $N_\textrm{pars}$ fit parameters.

\subsection{Fit Results}
To consider the possible interference between the $\chi_{c0}$ and non-resonant $\phiphi$ events, the latter are decomposed into their $0^+$ and $0^{-}$ components. Since the $\chi_{c1}$ and $\chi_{c2}$ are quite narrow, their interference with a non-resonant contribution is neglected. Therefore, only the possible interference between the $\chi_{c0}$ and non-resonant contributions is considered. The mass and width of the $\chi_{c0}$ are free parameters in the fit, and they are determined to be 3415.42 ${\rm MeV}$/$c^{2}$ and 11.4 $\rm{MeV}$ respectively, which are greater than the world average values (3414.71 $\pm$ 0.30 $\rm{MeV}$/$c^{2}$ and 10.8 $\pm$ 0.6 $\rm{MeV}$) from Ref.~\cite{rf12}. Meanwhile, the masses and widths of the $\chi_{c1}$ and $\chi_{c2}$ are fixed to the world average values, namely $M_{\chi_{c1}}=3510.67~{\rm MeV}/c^{2}, \Gamma_{\chi_{c1}}=0.84~{\rm MeV}$, $M_{\chi_{c2}}= 3556.17~{\rm MeV}/c^{2}$, and $\Gamma_{\chi_{c2}}=1.97~\rm{MeV}$. 

\begin{figure}[htbp]
\begin{center}
\mbox{
   \begin{overpic}[scale=0.35]{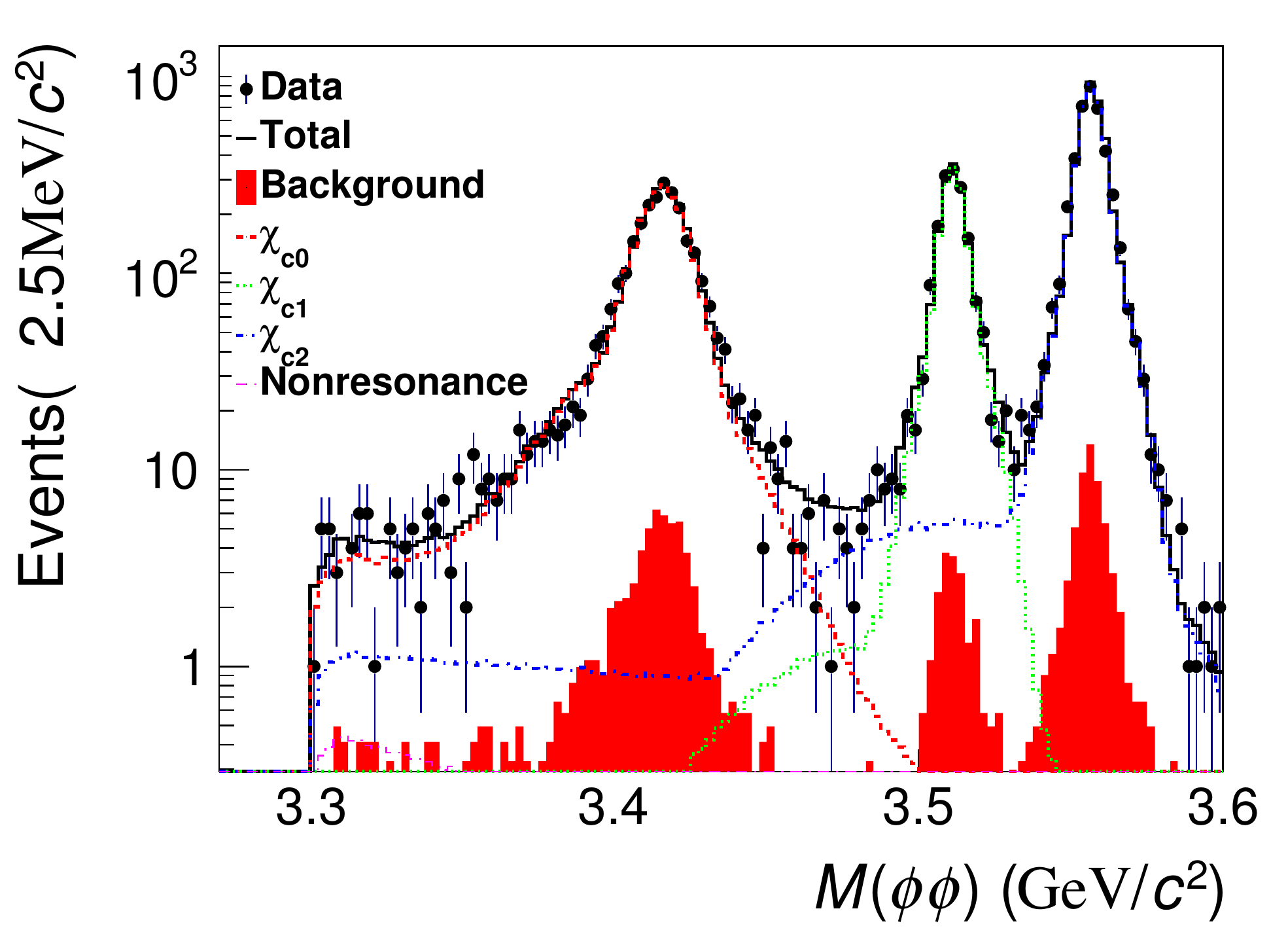}\end{overpic}
   \begin{overpic}[scale=0.35]{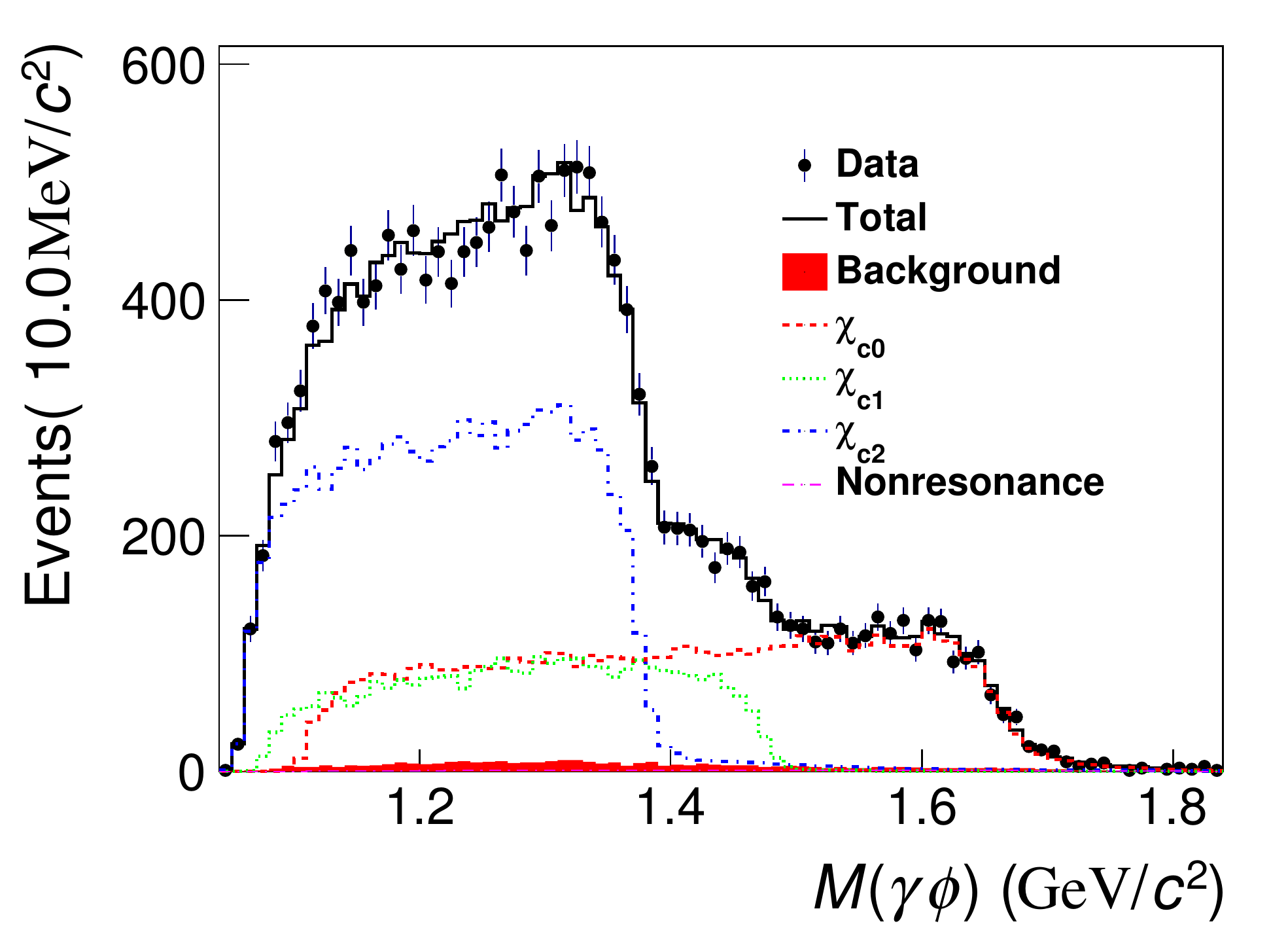}\end{overpic}
   }
\caption{Fit results of invariant mass distributions, $m_{\phi\phi}$ in the log version (left) and $m_{\gamma\phi}$ (right). The points with error bars represent data events. The black solid curve denotes the total fit result. The $m_{\gamma\phi}$ distribution has
two entries per event. Distributions of non-resonant events are almost invisible owing to the small contribution of this component.
   \label{fig fitchicj}}
\end{center}
\end{figure}

\begin{figure}[htbp]
\begin{center}
\mbox{
   \begin{overpic}[scale=0.37]{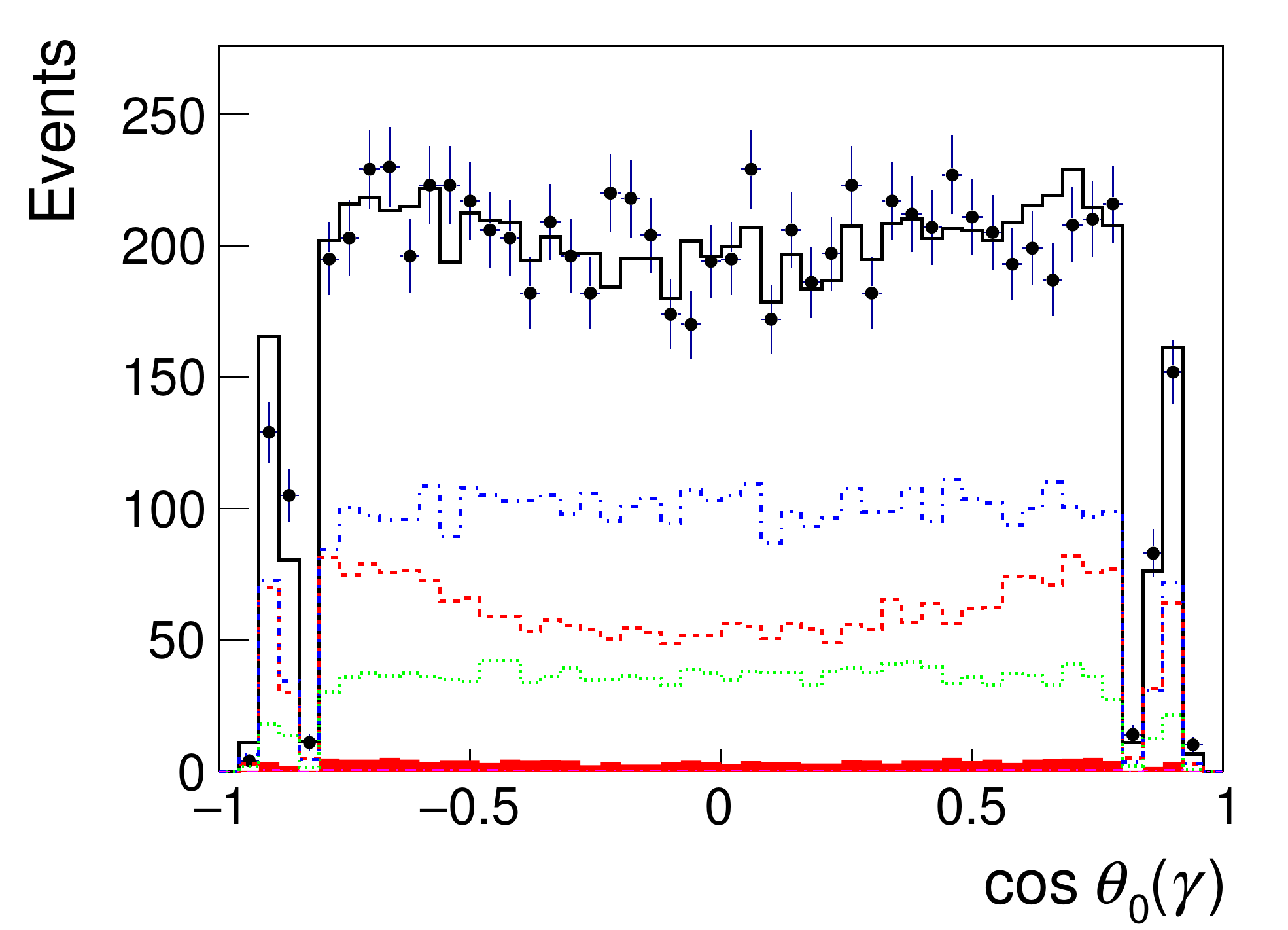}
   \end{overpic}
   \begin{overpic}[scale=0.37]{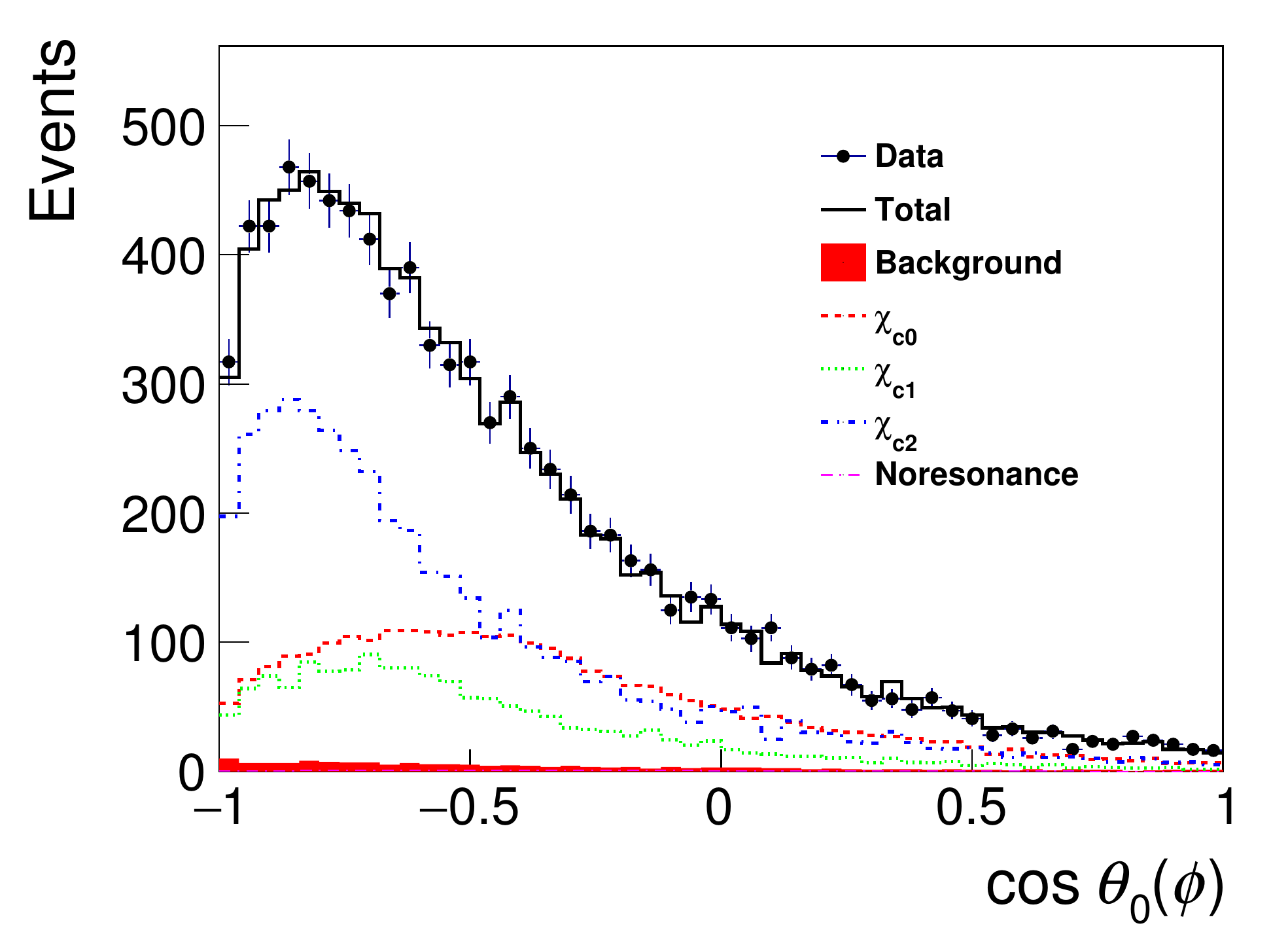}
   \end{overpic}
   }
\caption{Fit results of helicity angular distributions for the photon (left) and the $\phi$ meson (right). The points with error bars represent data events; The black solid curve denotes the total fit result. The red dashed, green dotted, and blue dashed-dotted curves represent the $\chi_{c0},~\chi_{c1}$ and $\chi_{c2}$ components, respectively.
\label{fig angs}}
\end{center}
\end{figure}

The $\phi\phi$ and $\gamma\phi$ invariant mass, as well as helicity angular distributions of $\gamma$ and $\phi$, are shown in Fig.~\ref{fig fitchicj} and Fig.~\ref{fig angs}, respectively.

With the fitted parameters, the fractions of signal yields in Eq.~(\ref{yieldsFormula}) are determined to be $(31.79\pm0.99)$\%, $(17.99\pm0.53)$\%, and $(49.98\pm1.09)$\% for the decay $\psip\to\gamma\chi_{cJ}$ followed by $\chi_{cJ}\to \phiphi~(J=0,1,2)$, respectively.
The ratio of amplitude moduli for $\chi_{c0}$ $\rightarrow$ $\phi \phi$ is determined to be
\begin{equation}
\begin{aligned}
		 x = \left|F_{1,1}^0\right|/\left|F_{0,0}^0\right| = 0.299\pm0.003.
\end{aligned}
\end{equation}

For $\chi_{c1}$ $\rightarrow$ $\phi \phi$, the ratios of amplitude moduli are determined to be
\begin{equation}
u_1=|F^1_{1,0}/F^1_{0,1}|=1.05\pm 0.05 \text{~and~} u_2=|F^1_{1,1}/F^1_{1,0}|=0.07\pm0.04.
\end{equation}
These two ratios are well consistent with the expectation of identical particle symmetry and parity conservation in strong decays.

For $\chi_{c2}$ $\rightarrow$ $\phi \phi$, the ratios of amplitude moduli are calculated to be
\begin{equation}
\begin{aligned}
\omega_{1} &= \left|F_{0,1}^2\right|/\left|F_{0,0}^2\right|  = 1.265\pm0.054,\\
\omega_{2} &= \left|F_{1,-1}^2\right|/\left|F_{0,0}^2\right| = 1.450\pm0.097,\\
\omega_{4} &= \left|F_{1,1}^2\right|/\left|F_{0,0}^2\right|  = 0.808\pm0.051,
\end{aligned}
\end{equation}
where the uncertainties are statistical only.

The branching fractions for $\chicJ\to\phiphi$ are determined by
\begin{equation}
Br[\chi_{cJ} \to \phi\phi]=
                        \frac{N}{\epsilon \ast N_{\psi}(3686) \ast Br[\psi(3686) \to \gamma\chi_{cj}] \ast Br^2[\phi \to K^+K^-]},
\end{equation}
where $N$ is $\chicJ$ signal yield which has been defined in Eq.~(\ref{yieldsFormula}), $N_{\psi(3686)}$ is the number of $\psi(3686)$ events~\cite{numofpsi}, $\epsilon$ is efficiency, and Br[$\psi(2S) \to \gamma\chi_{cJ}$($\phi\to K^{+}K^{-}$)] is the branching fractions for $\psip \to  \gamma\chi_{cJ}$($\phi\to K^{+}K^{-}$) quoted from Ref.~\cite{rf12}. The branching fractions for $\chicJ\to\phiphi$ with statistical uncertainties only and the parameters involved in calculation are summarized in Table~\ref{tab_br}.

\begin{table}[H]
\caption{\label{tab_br} Summary of the branching fractions for $\chicJ\to\phiphi$ and the parameters included in calculation.}
\begin{center}
\begin{tabular}{c|c|c|c}
 \toprule
Channel                                & $\chi_{c0} \to \phi \phi$   &   $\chi_{c1} \to \phi \phi$   &  $\chi_{c2} \to \phi \phi$ \\
 \midrule
$N$                                                          & 2701$\pm$84      & 1529$\pm$45      & 4247$\pm$93                              \\
$\epsilon$(\%)                                          & 29.73$\pm$0.10 & 34.09$\pm$0.10   & 32.60$\pm$0.10                                  \\
$N_{\psi (3686)}(\times 10^{6})$              &447.9$\pm$2.9   &447.9$\pm$2.9       &447.9$\pm$2.9 \\
$B(\psi(3686) \to \gamma\chi_{cj})$ (\%) &10.0$\pm$0.3     &9.6$\pm$0.3           &9.1$\pm$0.3 \\
$B(\phi \to K^+K^-)$ (\%)                         &48.9$\pm$0.5     &48.9$\pm$0.5         &48.9$\pm$0.5 \\
$B(\chi_{cJ}\to\phi\phi)(\times10^{-4})$   &8.48$\pm$0.26   &4.36$\pm$0.13       &13.36$\pm$0.29 \\
 \bottomrule
\end{tabular}
\end{center}
\end{table}

%========================================================
\section{\label{sec:level6}SYSTEMATIC UNCERTAINTIES}
The systematic uncertainties of helicity amplitude measurements are associated with the value of Blatt-Weisskopf barrier factor $d$, the tracking efficiency, the photon detection, the kinematic fit, the background estimation, and the mass resolution. Additionally, the number of $\psip$ events represents a source of uncertainty for the branching fraction determination.

\begin{itemize}

\item Branching fraction of intermediate states:\\
Branching fraction uncertainties for the decays $\BR(\psip \to  \gamma\chi_{c0})$, $\BR(\psip \to  \gamma\chi_{c1})$, $\BR(\psip \to  \gamma\chi_{c2})$, and $\BR(\phi \to K^{+}K^{-}$) are equal, respectively, to $3.0\%$, $3.1\%$, $3.3\%$, and $1.0\%$ Ref.~\cite{rf12}.

\item Blatt-Weisskopf barrier factor $d$\\
The systematic uncertainty arising from this term is determined via varying the factor $d$ between 1.5 and 4.5 $\rm{GeV}^{-1}$. The largest deviation from the nominal fit is determined to be 0.31$\%$, which is negligible.

\item Tracking efficiency and photon reconstruction:\\
    Tracking efficiencies for charged kaons are determined with a control sample of $J/\psi \to K^-K^{*+}$, $K^{*+} \to K^+\pi^0$ for $K^+$ and $J/\psi \to K^+K^{*-},K^{*-} \to K^-\pi^0$ for $K^-$, respectively~\cite{rf14}. The uncertainty due to the photon reconstruction is determined using a control sample of $J/\psi \to \rho^0\pi^0 \to \pi^+\pi^- \gamma\gamma$~\cite{rf17}. The ratio of efficiencies between data and MC simulation, $r(p_t,\cos\theta)={\epsilon_{\rm data}(p_t,\cos\theta)/ \epsilon_{\rm MC}(p_t,\cos\theta)}$, is taken as a factor to weight the squared amplitude of MC events to match that of data events dependent on the transverse momentum and polar angle of the track or shower. The difference of results with or without the weighting factor is taken as the systematic uncertainty.

\item Kinematic fit:\\
The systematic uncertainty from the kinematic fit is caused by the discrepancy between data and MC simulation in shower parameters for photons and track parameters for charged tracks. The simulation of photons has been investigated in Ref.~\cite{rf17}, which shows good agreement between data and MC simulation. For the charged tracks, the track helix parameter correction method ~\cite{rf1, rf13} is used. The systematic uncertainty is determined by replacing the MC sample with the helix-parameter-corrected ones when calculating the $\sigma_{\rm{MC}}$.

\item $\chi_{c1}$ and $\chi_{c2}$ masses and widths:\\
Alternative fits are performed where the resonance parameters of $\chi_{c1}$ and $\chi_{c2}$ are fixed to the values sampled within one standard deviation of the PDG values~\cite{rf12}. The resultant differences from the nominal fit are taken as the systematic uncertainties.

\item Mass resolution:\\
The uncertainty of the mass resolution is determined by replacing the resolution functions obtained from the signal MC samples with the new one obtained from the control samples in data.

\item Background estimation:\\
The number of background events, estimated to be $N_\text{bkg}=166\pm13$, is subtracted in the fit. The systematic uncertainty due to the background fluctuation is estimated by assuming a Gaussian distribution, $G(N_\text{bkg},\delta N_\text{bkg})$, whose width is $\delta N_\text{bkg}=13$, which is taken as a new weighting factor propagated to amplitude model by modifying the log-likelihood function as $G(N_\text{bkg},\delta N_\text{bkg})\ln\mathcal{L}_{\text{bkg}}$.

\item Number of $\psip$ events:\\
The number of $\psip$ events is measured by studying inclusive hadronic decays. The uncertainty is about 0.6\% \cite{numofpsi}.

\item
The possible correlations among the tracking efficiency, photon reconstruction, kinematic fit calculation, $\chi_{c1}$ and $\chi_{c2}$ masses and widths, mass resolution discrepancy and background estimation are considered in alternative fits. The normalization factor $\sigma_\text{MC}$ in Eq.~(\ref{normfct}) is modified as
\begin{equation}
\sigma_\text{MC}={1\over N_\text{MC}}\sum_{i=1}^{N_\text{MC}}\overline{\sum}|\mathcal{M}(\chicz)+\mathcal{M}(\chico)+\mathcal{M}(\chict)+\mathcal{M}(NR)|^2T_\text{corr},
\end{equation}
where the factor $T_\text{corr}$ is defined as the product of the correction factors for the $\gamma$ and $2(K^+K^-)$ tracking ratios, $r(p_t,\cos\theta)$, for a given event, and the masses and widths of $\chico$ and $\chict$ are sampled within one standard deviation of the PDG values~\cite{rf12}. The mass resolution function for $\chicz$ is replaced by the alternative one. The MC events are replaced by the helix-parameter-corrected ones. 
Then alternative fits are performed to obtain the $\chicJ$ helicity amplitude ratios, the mass and width of $\chi_{c0}$, and the signal yields for each resonance. The differences with individual nominal results are taken as the correlated systematic uncertainties of the amplitude analysis, as given in Table~\ref{tab6}. It is notable that the determined uncertainty for the ratio of amplitude moduli of the helicity conserved amplitudes ($\left|F_{1,1}^2\right|/\left|F_{0,0}^2\right|$) is smaller than those of the other helicity violated amplitudes. This is due to the fact that in the helicity conserving amplitude, the kinematic fit and other systematic errors have opposite signs and cancel each other, while in the helicity violating amplitude, the signs are the same and the errors are constructive.

\end{itemize}
  \begin{table}[H]
    \caption{\label{tab6} Summary of systematic uncertainties. The correlated sources include the tracking efficiency, photon reconstruction, kinematic fit, backgrounds, $\chi_{c1}$ and $\chi_{c2}$ masses and widths and mass resolution. $\BR(\chicj)$ denotes the branching fraction for $\chicj\to\ff$. The "---" means that this term is not available in the corresponding measurement. }.
    \begin{center}
    \begin{tabular}{lcccc}
      \toprule
      Source & $\BR(\chi_{c0}) $   &   $\BR(\chi_{c1}) $   &  $\BR(\chi_{c2}) $ & \\
      \midrule
      $N_{\psip}$ & 0.6\% &  0.6\% & 0.6\% &\\
      $\BR(\psip \to \gamma\chi_{cJ})$  & 3.0\% & 3.1\%  & 3.3\%  & \\
      $\BR(\phi \to K^+K^-)$      &1.0\%  & 1.0\%  & 1.0\% &\\
      Correlated sources & 0.4\% & 2.3\% & 1.1\%& \\
      Total & 3.2\% & 4.0\% & 3.7\%&\\\hline
      Sources & ${|F_{1,1}^0|\over|F_{0,0}^0|}$ & ${|F_{0,1}^2|\over |F_{0,0}^2|}$ & ${|F_{1,-1}^2|\over|F_{0,0}^2|}$ & ${|F_{1,1}^2|\over|F_{0,0}^2|}$ \\\hline
      $N_{\psip}$ &---&---&---&---\\
      $\BR(\psip \to \gamma\chi_{cJ})$  &---&---&---&--- \\
      $\BR(\phi \to K^+K^-)$   &---&---&---&---\\
      Correlated sources &6.4\%&6.2\%&7.1\%&1.1\%\\
      Total &6.4\%&6.2\%&7.1\%&1.1\%\\
      \bottomrule
      \end{tabular}

    \end{center}
  \end{table}

\section{\label{sec:level8}CONCLUSION AND DISCUSSION}
Using (447.9 $\pm$ 2.3) million $\psip$ events collected with the BESIII detector, the helicity amplitudes for the decays $\psip\to\gamma\chicJ,\chicJ\to\phi\phi$, and $\phi\to\kk$ are studied. 

The branching fractions for $\chicJ\to\phiphi$ are measured to be
\begin{eqnarray}
\BR(\chi_{c0} \rightarrow \phi \phi)&=& (8.48\pm0.26\pm0.27)\times10^{-4} ,\nonumber \\
\BR(\chi_{c1} \rightarrow \phi \phi)&=& (4.36\pm0.13\pm0.18)\times10^{-4} , \\
\BR(\chi_{c2} \rightarrow \phi \phi)&=& (13.36\pm0.29\pm0.49)\times10^{-4}\nonumber,
\end{eqnarray}
where the first uncertainties are statistical and the second systematic. Comparing these results with BESIII previous measurement~\cite{rf14} and PDG values~\cite{rf12}, as reflected in the Table \ref{tab BFsum}, the precision is improved by a factor of about $2$, but the values are greater.

\begin{table}[htbp]
\begin{center}
\caption{ \label{tab BFsum} \footnotesize Comparsion of measured branching fractions (BF).}
%\vskip 0.1cm
\begin{normalsize}
%\begin{tabular}{ m{4cm}<{\centering} | m{3.5cm}<{\centering}  | m{3.5cm}<{\centering}  | m{3.4cm}<{\centering}  }
\begin{tabular}{ c|c|c|c }
\hline
\hline
Decay Mode & BF(2011 BESIII)~\cite{rf14} & BF(this work)& BF(PDG value)~\cite{rf12}\\
\hline
Br[$\chi_{c0}$ $\rightarrow$ $\phi \phi$]($\times$$10^{-4}$) &  7.8$\pm$0.4$\pm$0.8 & $8.48\pm0.26\pm0.27$ & 7.7$\pm$0.7\\
Br[$\chi_{c1}$ $\rightarrow$ $\phi \phi$]($\times$$10^{-4}$) &  4.1$\pm$0.3$\pm$0.5 & $4.36\pm0.13\pm0.18$ & 4.2$\pm$0.5\\
Br[$\chi_{c1}$ $\rightarrow$ $\phi \phi$]($\times$$10^{-4}$) &  10.7$\pm$0.4$\pm$1.2 & $13.36\pm0.29\pm0.49$ & 11.2$\pm$1.0\\
\hline\hline
\end{tabular}
\end{normalsize}
\end{center}
\end{table}

The ratios of the amplitude moduli are measured to be
\begin{eqnarray}
\left|F_{1,1}^0\right|/\left|F_{0,0}^0\right| &=& 0.299\pm0.003\pm0.019,
\end{eqnarray}
for $\chi_{c0}\to\ff$, and
\begin{eqnarray}
\left|F_{0,1}^2\right|/\left|F_{0,0}^2\right| &=& 1.265\pm0.054\pm0.079,\\
\left|F_{1,-1}^2\right|/\left|F_{0,0}^2\right| &=& 1.450\pm0.097\pm0.104,\\
\left|F_{1,1}^2\right|/\left|F_{0,0}^2\right| &=& 0.808\pm0.051\pm0.009,
\end{eqnarray}
for $\chi_{c2}\to\ff$, where the first and second uncertainties are statistical and systematic, respectively. Additionally, there is no evidence of identical particle symmetry breaking from the study of $\chi_{c1}\to\ff$.

\begin{figure}[H]
\begin{center}
\mbox{
   \begin{overpic}[scale=0.6]{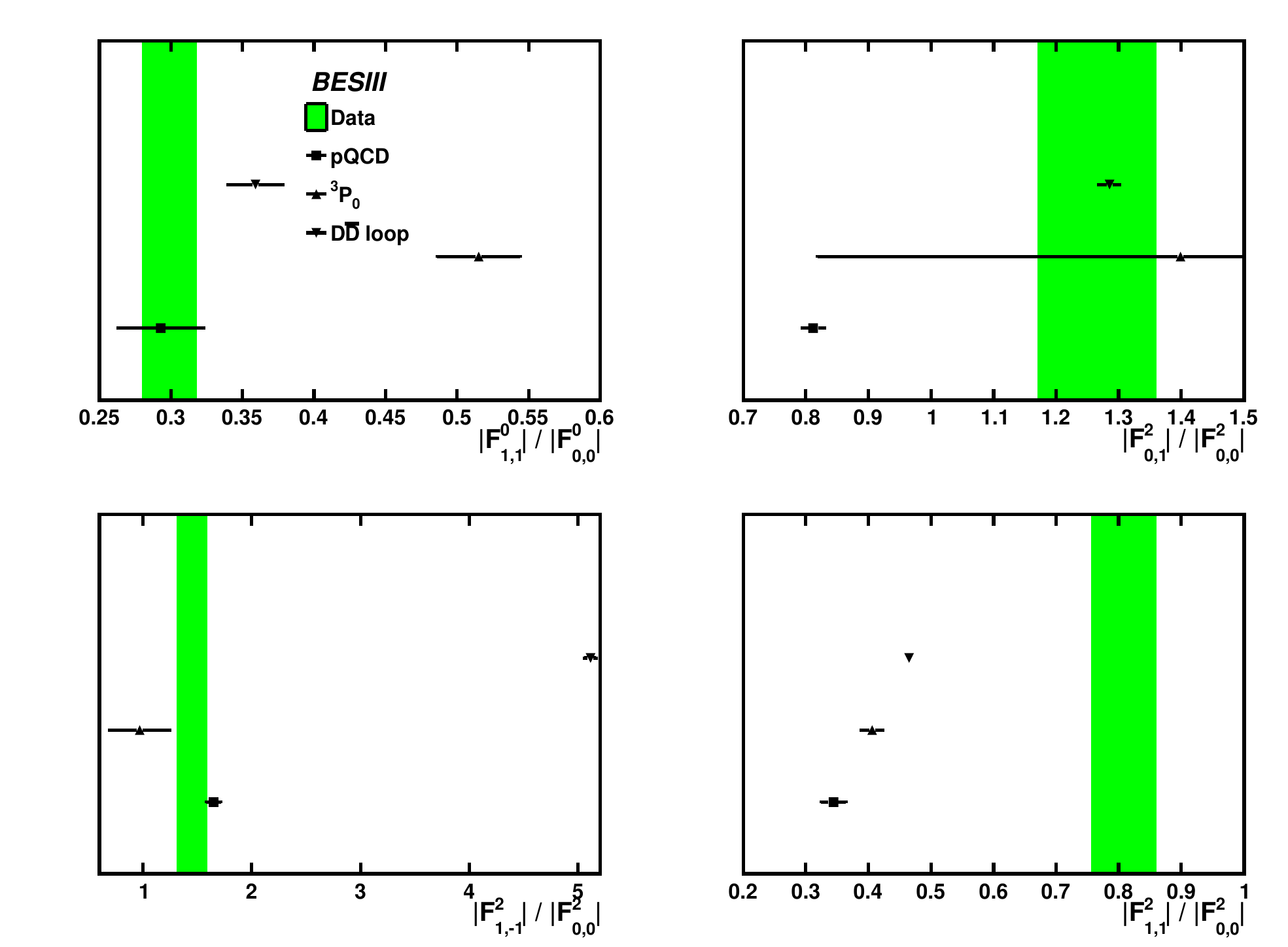} \end{overpic}
   }
\caption{Comparison of the measured amplitude ratios with the predicted ones from pQCD, the $^3P_0$ model and the $D\bar {D}$ loop model.
\label{cmptheo}}
\end{center}
\end{figure}

Figure~\ref{cmptheo} shows a comparison of the measured amplitude ratios to the corresponding theoretical predictions. The measured ratio of amplitude moduli for the $\chi_{c0}$ is consistent with the pQCD prediction of Ref.~\cite{rf3}, since two independent helicity amplitudes of the $\chi_{c0}\to\phiphi$ decay,  $F_{1,1}^0$ and $F_{0,0}^0$,  follow the helicity selection rule. For the $\chi_{c2}$ decay, the measured ratios of amplitude moduli deviate from the pQCD~\cite{rf3}, $^3P_0$~\cite{rf16} and $D\bar {D}$ loop~\cite{Huang:2021kfm} predictions with $\chi^2/\mathrm{ndf}=23.2$, $23.8$, and $155.2$, respectively. The $D\bar D$ loop model can be ruled out due to the large deviation. However, the predictions of other models also differ from the experimental results. In short, all of the above theories use some of the input from the experimental results, thus this measurement can provide more constraints for further developing the models. It could also be a basis for the measurement in the future, as 2.7 billion $\psi(3686)$ events have been accumulated in BESIII~\cite{white-paper}.

%\section{\label{sec:level9} ACKNOWLEDGEMENTS }
~\\\\
\noindent ACKNOWLEDGEMENTS\\\\
The BESIII collaboration thanks the staff of BEPCII and the IHEP computing center for their strong support. This work is supported in part by National Key R$\&$D Program of China under Contracts Nos. 2020YFA0406300, 2020YFA0406400; National Natural Science Foundation of China (NSFC) under Contracts Nos. 11875262, 12175244, 11635010, 11735014, 11835012, 11935015, 11935016, 11935018, 11961141012, 12022510, 12025502, 12035009, 12035013, 12192260, 12192261, 12192262, 12192263, 12192264, 12192265; the Chinese Academy of Sciences (CAS) Large-Scale Scientific Facility Program; Joint Large-Scale Scientific Facility Funds of the NSFC and CAS under Contract No. U1832207, U2032110; CAS Key Research Program of Frontier Sciences under Contract No. QYZDJ-SSW-SLH040; 100 Talents Program of CAS; The Institute of Nuclear and Particle Physics (INPAC) and Shanghai Key Laboratory for Particle Physics and Cosmology; ERC under Contract No. 758462; European Union's Horizon 2020 research and innovation programme under Marie Sklodowska-Curie grant agreement under Contract No. 894790; German Research Foundation DFG under Contracts Nos. 443159800, Collaborative Research Center CRC 1044, GRK 2149; Istituto Nazionale di Fisica Nucleare, Italy; Ministry of Development of Turkey under Contract No. DPT2006K-120470; National Science and Technology fund; National Science Research and Innovation Fund (NSRF) via the Program Management Unit for Human Resources $\&$ Institutional Development, Research and Innovation under Contract No. B16F640076; STFC (United Kingdom); Suranaree University of Technology (SUT), Thailand Science Research and Innovation (TSRI), and National Science Research and Innovation Fund (NSRF) under Contract No. 160355; The Royal Society, UK under Contracts Nos. DH140054, DH160214; The Swedish Research Council; U. S. Department of Energy under Contract No. DE-FG02-05ER41374

\bibliographystyle{jhep}
\bibliography{./bib/ref.bib}

\section{\label{sec:level9}THE BESIII COLLABORATION}
%\author{Author list}
\begin{small}
\begin{center}
M.~Ablikim$^{1}$, M.~N.~Achasov$^{11,b}$, P.~Adlarson$^{70}$, M.~Albrecht$^{4}$, R.~Aliberti$^{31}$, A.~Amoroso$^{69A,69C}$, M.~R.~An$^{35}$, Q.~An$^{66,53}$, X.~H.~Bai$^{61}$, Y.~Bai$^{52}$
% ...
\\
\vspace{0.2cm}
(BESIII Collaboration)\\
\vspace{0.2cm} {\it
$^{1}$ Institute of High Energy Physics, Beijing 100049, People's Republic of China\\
% ...
$^{4}$ Bochum Ruhr-University, D-44780 Bochum, Germany\\
% ...
$^{11}$ G.I. Budker Institute of Nuclear Physics SB RAS (BINP), Novosibirsk 630090, Russia\\
% ...
$^{31}$ Johannes Gutenberg University of Mainz, Johann-Joachim-Becher-Weg 45, D-55099 Mainz, Germany\\
% ...
$^{35}$ Liaoning Normal University, Dalian 116029, People's Republic of China\\
% ...
$^{52}$ Southeast University, Nanjing 211100, People's Republic of China\\
$^{53}$ State Key Laboratory of Particle Detection and Electronics, Beijing 100049, Hefei 230026, People's Republic of China\\
% ...
$^{61}$ University of Jinan, Jinan 250022, People's Republic of China\\
% ...
$^{66}$ University of Science and Technology of China, Hefei 230026, People's Republic of China\\
% ...
$^{69}$ University of Turin and INFN, (A)University of Turin, I-10125, Turin, Italy; (B)University of Eastern Piedmont, I-15121, Alessandria, Italy; (C)INFN, I-10125, Turin, Italy\\
$^{70}$ Uppsala University, Box 516, SE-75120 Uppsala, Sweden\\
% ...
$^{b}$ Also at the Novosibirsk State University, Novosibirsk, 630090, Russia\\
% ...
}\end{center}

\vspace{0.4cm}
\end{small}

\end{document}